# Conceptual Design of Mars Lander with Novel Impact Intriguing System

Malaya Kumar Biswal M[*] and Ramesh Naidu Annavarapu[‡]

*Department of Physics, School of Physical Chemical and Applied Sciences,*

*Pondicherry University, Kalapet, Puducherry, India – 605 014*

**Landing robotic spacecrafts and humans on Mars has become one of the inevitable technological necessities for humans. Effectuating perfect Mars expedition requires landing of enormous cargoes, crewed modules, ascent vehicles, and scientific laboratories. Crash-landing destructs landing modules due to a lack of adequate impact absorbers. Moreover, the existence of deformable shock absorbers like Aluminium honeycomb and carbon fibers within landing gears are defeasible for large-scale mass and crewed landing. Further in the EDL scenario, switching of events within a limited span of 5 to 8 minutes appears to be the most challenging task for landers. Scrutinizing these concerns, we propose a novel impact intriguing system that will be more achievable for extensive landing missions. This paper represents the conceptual design of Mars lander and we expect that subject to any obstruction in EDL sequence, this mechanical system will endorse soft-landing in forthcoming missions. Hence our ultimate aim is to protect lander modules and their instruments over the defective landing.**

## I. Nomenclature

| | | |
|---|---|---|
| *EDL* | = | Entry, Descent and Landing |
| *ESA* | = | European Space Agency |
| *HIAD* | = | Hypersonic Inflatable Aerodynamics Decelerator |
| *JAXA* | = | Japan Aerospace Exploration Agency |
| *LEO* | = | Low Earth Orbit |
| *MPF* | = | Mars Pathfinder |
| *MPL* | = | Mars Polar Lander |
| *NASA* | = | National Aeronautics Space Administration |
| *ROSCOSMOS* | = | Russian State Corporation for Space Activities |

## II. Introduction

Making humans a multi-planetary species has become a technological progression of mankind. To meet this goal, we need to land a large number of scientific cargoes and crewed landers on Mars. But the incidences of crash landing intimidated us to anticipate effective human landing. An earlier mission such as Viking and Venera landers to Mars and Venus used Aluminium foam and honeycomb as an impact attenuator as it possesses high impact absorption efficiency. Alike US landers Soviet landers also used foam plastic as their primary impact armor. Despite good shock absorbers, some landers fail to perform ground touchdown due to inappropriate EDL performance within a limited EDL period. To address this technological challenge, we propose a novel impact intriguing system to achieve soft-landing excluding any obstruction in EDL performance.

[*] Graduate Researcher, Department of Physics, Pondicherry University, India. Member of Indian Science Congress Association, Student Member AIAA, malaykumar1997@gmail.com; mkumar97.res@pondiuni.edu.in

[‡] Associate Professor, Department of Physics, School of Physical, Chemical, and Applied Sciences, Pondicherry University, Puducherry, India. rameshnaidu.phy@pondiuni.edu.in; arameshnaidu@gmail.com.

## III. History of Earlier Lander Missions

### A. Earlier Mars Lander Missions

The pioneer attempt to touch the red planet begins with the historic launch of lander 2MV-3 No.1 onboard Molniya that endured failure and destroyed in LEO due to launch vehicle concerns. Besides this misfortune, Soviet Union re-attempted Mars 2 & 3 landers whereas Mars 2 remained ineffective and Mars 3 lasted for 20 seconds of data transmission after landing. Alike Mars 2 & 3, Russian again re-attempted to land on Mars with Mars 5, 6 &7, but entire effort lasted disappointment as Mars 6 malfunctioned and Mars 7 missed the planet before atmospheric entry. In 1975, the United States first landing attempt with twin Viking landers (Viking 1 & 2) performed extraordinary landing with successful mission accomplishment. Aspiring to explore Martian moons, Soviet Union strived dual Phobos mission each comprising of orbiter and lander, but both missions encountered radio communication issues. The advancement of new EDL technology (i.e., parachute deployment and airbag landing system) made the US achieve second landing (Mars Pathfinder-1966) which deployed a mini-rover and Mars Phoenix lander in 2007. Subsequent lander missions like the collapse of Russian's Mars 96 caused by upper-stage booster concerns, Mars Polar lander loss as a consequence of communication issues.

In the controversy of the Mars expedition, ESA (Europe) comes into execution with its first landing endeavor (Beagle 2 lander) that enforced effective landing but unintentionally obscured communication. Their effort pursued with second landing attempt (Schiaparelli EDM) which achieved successful hypersonic atmospheric entry into the Martian atmosphere but shortly, premature deployment of parachute made it crash on the surface. Having upgraded Viking era technology, the United States successfully landed InSight Mars lander in November 2018 and it continues its surface operation on Mars. Some of the failed landers are listed below in table-1.

**Table-1 Crashed Landers**

| Lander | Issue | Results |
|---|---|---|
| **Mars 2 Lander** | Improper atmospheric entry | Crashed on the surface |
| **Mars Polar Lander** | Lost communication before touchdown | Failed |
| **Deep Space 2 Penetrator** | Lost with Mars Polar Lander | Failed |
| **Schiaparelli EDM** | Faulty gyroscope damaged guidance control system | Crashed on the surface |

## IV. Landing Challenges and Prospect

For dynamic landing, the lander has to confront distinct challenges such as the thickness of the Mars atmosphere, low surface elevation, limited EDL period, and surface hazards which include distribution of large seized rocks, valleys, terrains and robust dust storms. Especially for legged landers have rock hazards as their largest challenge. Crushable bumper landers having aluminum honeycomb, foam plastic, airbags, and crushable carbon fibers. And they don't have to contemplate surface hazards but competent aerodynamic deceleration due to mass constraints (i.e., 0.6 ton). In the conception of future manned lander missions, it is quite requisite to ground large masses, and on other hand factors such as the diameter of aeroshell & parachute and ballistic coefficient are restrained. To execute effective human class Mars mission with extended traditional EDL technology mass constraints extended from 0.6 ton to 25 tons and it may yield more impact force upon landing. Hence our novel impact intriguer may sustain effective touchdown.

**Table-2 Past Landers and their Impact Attenuators**

| | Foam Plastic | Aluminium Honeycomb | Air-Bags | Crushable Materials | Inflatable Ballute |
|---|---|---|---|---|---|
| **Landers** | Mars 2,3,6,7<br>Phobos 1,2 | Viking 1,2 | MPF,<br>Mars96, Beagle 2 | Phoenix, Schiaparelli,<br>Insight | Deep Space 2 |
| **Mass (kg)** | 358 – 635 | 423 – 590 | 30 – 360 | 280 – 360 | 4.5 |

## V. Design Methodology and Working

### A. Design and Modelling of Lander Base

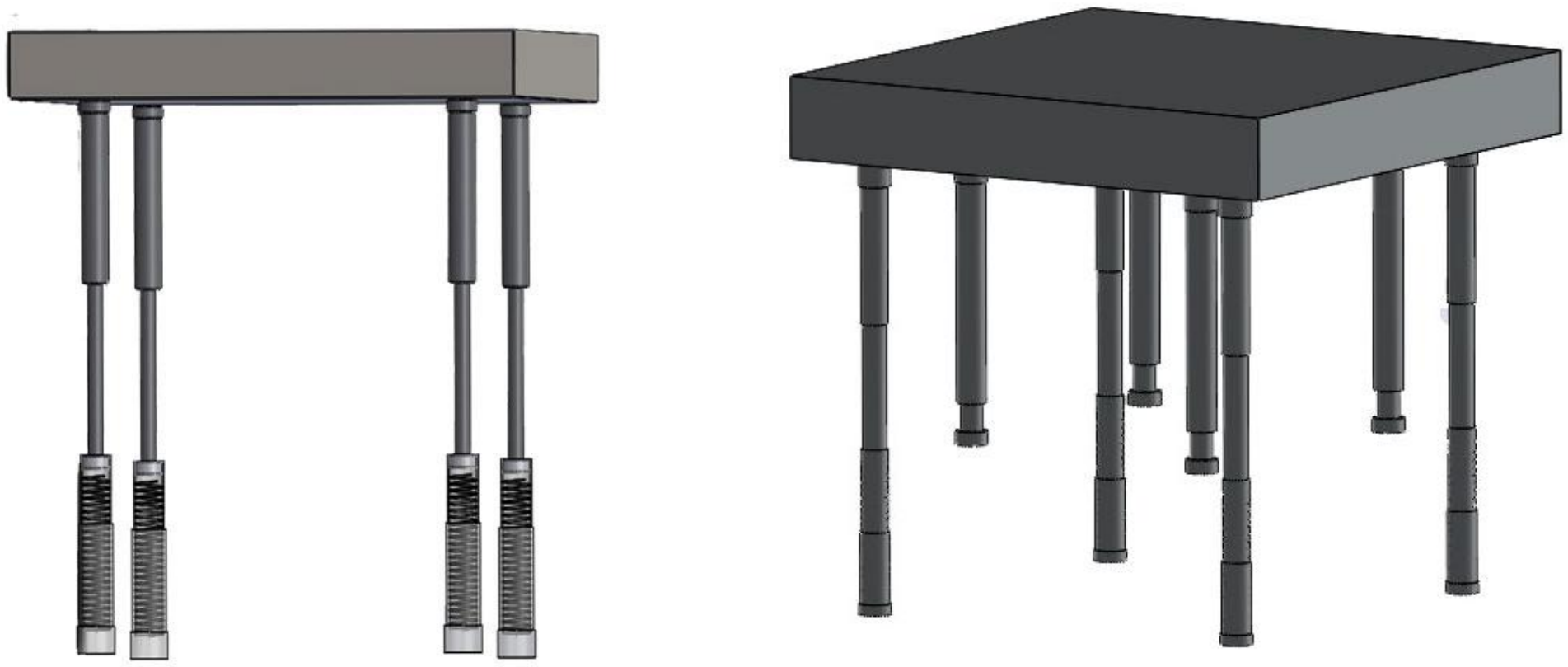

**Fig 1 Primary and Secondary Structure of Lander Base**

The conceptual design of Mars lander has a lander base comprises of eight landing legs in both inner square lines (four hydraulic legs) and outer square lines (four novel impact landing legs). Each leg comprising of primary attenuator with a spring-damper and secondary attenuator with a hydraulic damper. The primary attenuator incorporates a spring holder and a helical compression spring as well as a secondary attenuator incorporates a hydraulic cylinder and a piston rod. Moreover, the hydraulic cylinder divides into the upper and lower compartment where the upper compartment is loaded with compressed oil and air that are the best adoptive method for landers. The internal portion of the upper compartment has a mounted control valve to release the hydraulic fluid and a pressure sensor to indicate pressure range to control valve. The function of the sensor is to signal the sensor to open and gradually exempt hydraulic fluid when the pressure exceeds beyond a predetermined level. Conclusively total setup is confined within a retractable cylinder which completes the entire lander leg setup. In addition to this, we excogitated to install another four hydraulic cylinders which are supposed to work when the pressure of the secondary unit exceeds beyond the predetermined level.

### B. Modelling of Lander Leg

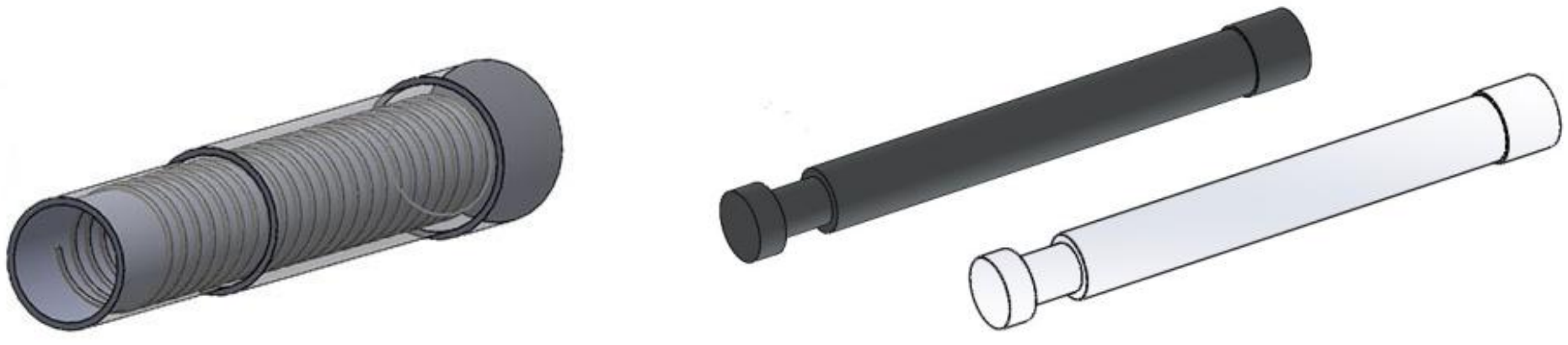

**Fig 2 Primary and Secondary Unit of Lander Leg**

**Primary Unit:** The primary unit has high tensile Si-Cr helical compression spring as a principal shock absorber. The spring is wrapped inside a retractable cylinder coupled structure of appropriate diameter. The geometry of spring relies on spring constant that ultimately depends on the lander mass, the vertical distance from where the lander gets dropped, and the terminal velocity. The logic behind the use of spring-damper as principal shock attenuator is due to the expeditious nature of impact absorption exerted during ground touchdown.

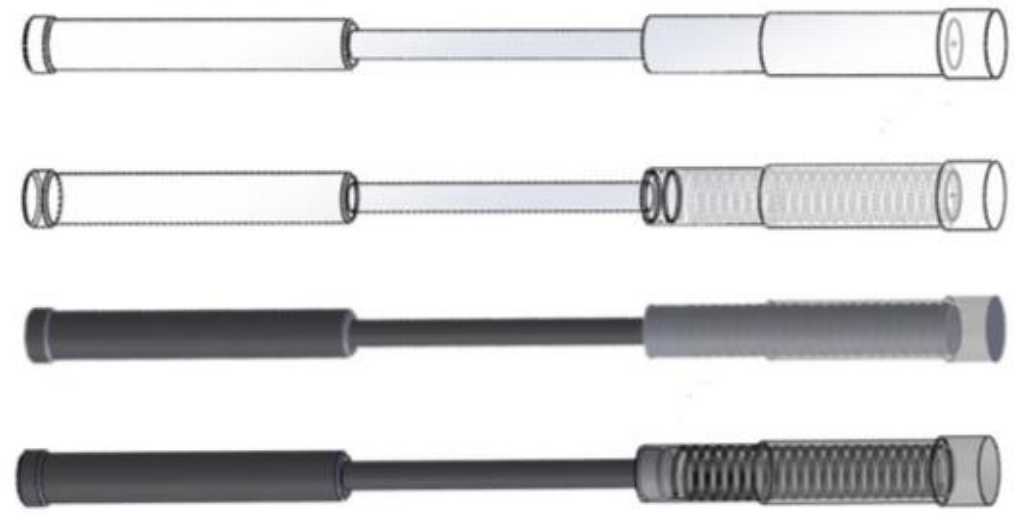

**Fig 3 Structure of Lander Legs**

**Secondary Unit:** The primary unit is integrated with a hydraulic damper as a secondary unit. The hydraulic systems absorb the shock with the help of a hydraulic accumulator attached to that system. Here, the basis of utilizing hydraulic damper as an auxiliary shock absorber is due to the nature of rapid energy absorption (exerted by the spring) and moderate energy dissipation.

### B. Working

During the interim of the landing phase, after the halt of retrorockets, the lander gets dropped from a certain altitude, and the lander contacts the ground from a vertical direction. It gets subjected to vertical impact now the spring damper inside primary attenuator starts working and gets pressed downwards, the maximum impact gets absorbed by the spring and stores in the form of mechanical energy. After a short layoff, it transfers the same energy to secondary attenuator obeying newton's second law.

Secondary attenuator encounters maximum impact which in turn makes its piston to move upwards so that the hydraulic damper can obsess and dissipate maximal impact. Relatively, some small amount of hydraulic fluid will move from upper to the lower compartment. In case of excess hydraulic pressure, a pressure sensor installed inside the cylindrical chamber will signal the control valve to exempt pressurized fluid. As a consequence, the maximum impact gets attenuated by an additional four hydraulic legs and the lander may attain effective landing on Mars.

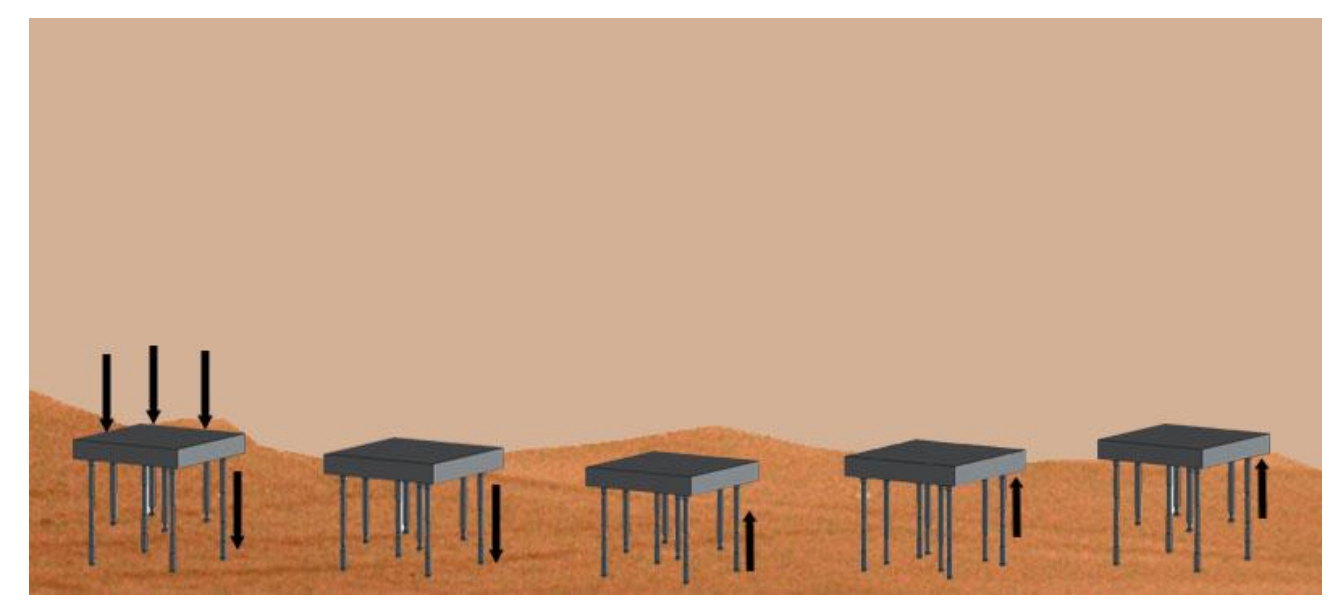

**Fig 4 Animated Image of Working of Lander**

## VI. Mathematical Formulation

Here the area of the outer four landing legs is smaller than the inner four landing legs ( $A_2 > A_1$ ) and this implies the force exerted on $F_2 > F_1$. So, the velocity of hydraulic fluid moving across A is $v_1$ and the velocity at B is $v_2$. And the ( $A_2 > A_1$ ) reflects that ($v_2 < v_1$). Hence from the equation of continuity, $a_1 v_1 = a_2 v_2$ where $av = constant$.

Further, the force acting on the fluid at A equals $P_1 a_1$ and the force acting on fluid at B equals $P_2 a_2$. Similarly, the work done on the fluid at A equals $P_1 a_1 \times v_1 = P_1 V$ and the work done by the liquid at B equals $P_2 a_2 \times v_2 = P_2 V$.

Evaluating the work done per second by the pressure energy relation which equals $P_1 V - P_2 V$. Additionally, the increase in kinetic energy over landing can be expressed as

$$\left[\frac{1}{2}mv_2^2 - \frac{1}{2}mv_1^2\right]$$

And the increase in potential energy can be expressed as

$$[mgh_2 - mgh_1]$$

Applying work-energy principle over work done, kinetic and potential energy, we get two force equations $F_1$ *and* $F_2$

$$P_1 V - P_2 V = (mgh_2 - mgh_1) + \left(\frac{1}{2}mv_2^2 - \frac{1}{2}mv_1^2\right)$$

$$P_1 V = P_2 V + mg\,(h_2 - h_1) + \frac{1}{2}m(v_2^2 - v_1^2)$$

$$\frac{F_1}{A_1}V = P_2 V + mg\,(h_2 - h_1) + \frac{1}{2}m(v_2^2 - v_1^2)$$

$$F_1 = \frac{P_2 V A_1}{V} + \frac{mg\,(h_2 - h_1)A_1}{V} + \frac{m(v_2^2 - v_1^2)A_1}{2V}$$

$$F_1 = P_2 A_1 + \frac{mA_1}{V}\left[g(h_2 - h_1) + \frac{1}{2}(v_2^2 - v_1^2)\right] \text{ and } F_2 = P_1 A_2 - \frac{mA_2}{V}\left[g(h_2 - h_1) + \frac{1}{2}(v_2^2 - v_1^2)\right]$$

The equation shows that the force exerted over $F_2$ is greater than $F_1$.

**Fig 5 Free Body Diagram of Leg Interlink**

## VII. EDL Performance and Applications

### A. Atmospheric Entry

The descent vehicle is excogitated to enter from Martian orbit to minimize the entry velocity rather than direct entry. It is the safest approach for manned and large scale landing. It eliminates difficulties such as dust storms that can be predicted in advance by various Mars orbiters and it will allow us for effective preparation for Mars atmospheric entry.

### B. Descent

The entry phase is followed by a descent phase, where the lander undergoes fast aerocapture by the rapid action of parachute deployment and Hypersonic Inflatable Aerodynamic Decelerator. Using the Viking technology NASA had landed the largest mass up to 900kg of Mars Science Laboratory. It used a 70° Spherical cone shell with a diameter of 4.5m. The need to increase the diameter of aeroshell to decrease the ballistic coefficient and faster aerocapture is limited due to the availability of launch vehicles. To rectify this situation Hypersonic Inflatable Aerodynamics Decelerators can be employed for faster aerocapture during the descent phase.

### C. Landing

After successful descent, the lander undergoes a landing phase where the novel legs will be deployed to rest the lander on the surface of Mars. Due to the effective design of the lander, the legs were in a deployed position and restricts the use of leg deployment events. Our novel EDL follows vertical landing for un-deployable legs. Now the lander gets successfully descended on the surface.

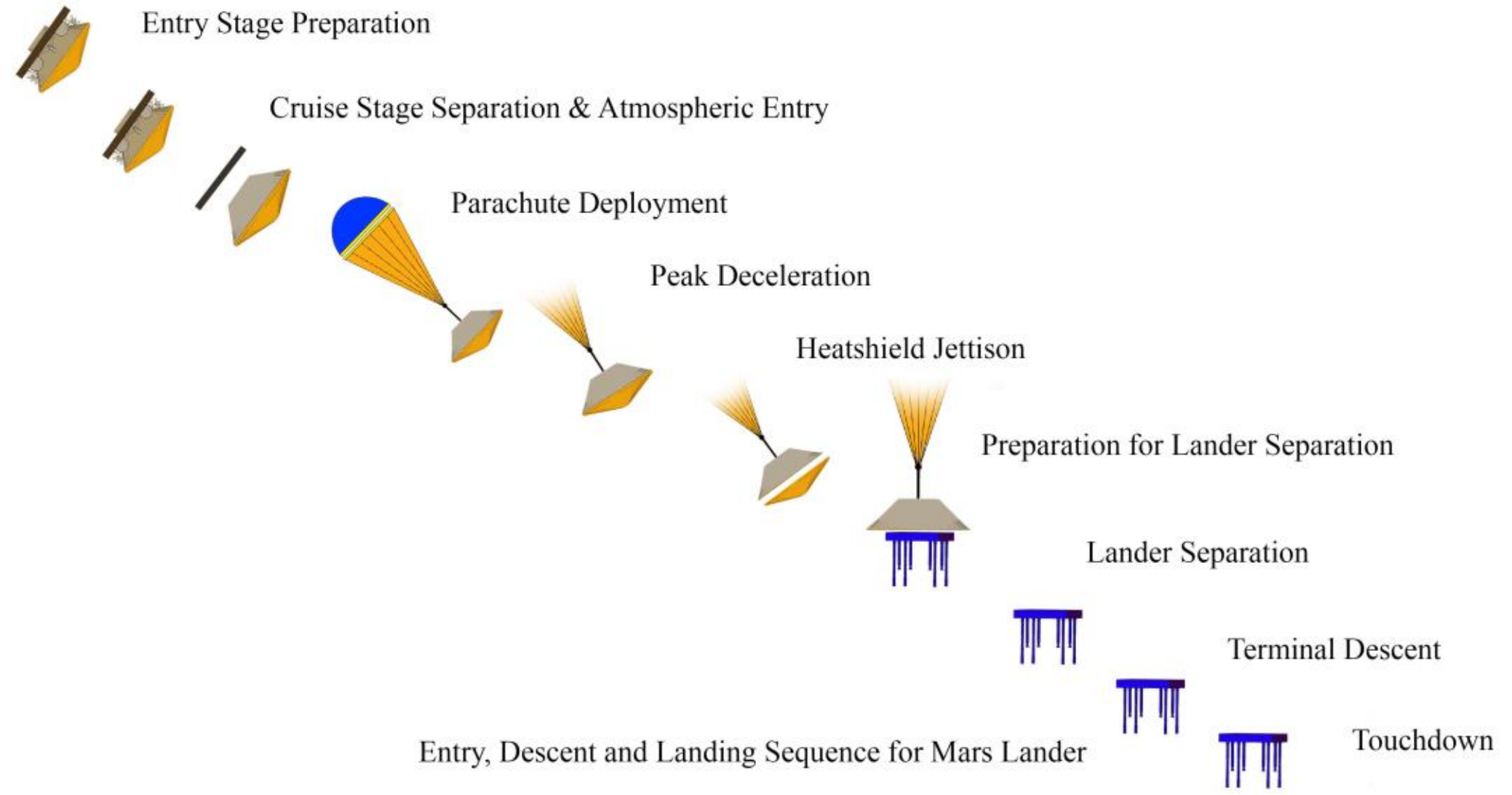

**Fig 6 EDL Map of Mars Lander**

### D. Applications

This novel method of impact attenuation system will enable safe and large scale mass landings on the surface of Mars. It was devoid of the use of expensive retropropulsive systems thereby promoting cost-efficient landing on Mars. It has very good shock absorption capacity and the shock absorption capacity of the novel leg varies according to the size, density of the hydraulic fluid, and the spring constant defined during fabrication. It will have excellent applications shortly for future human Mars exploration missions and large scale cargo missions. Some of the future proposals lander mission understudy is mentioned in the table-4, which expresses the future of this novel leg.

**Table-3 Comparative EDL Summary of Mars Landers**

| Parameters | Mars 2 | Mars 3 | Mars 6 | Mars 7 | Viking-1 | Viking-2 | Phobos 1 and 2 | MPF | Mars 96 | Mars 96 Penetrator | MPL | Beagle 2 | Deep 2 Space | Phoenix Lander | Schiaparelli EDM | InSight |
|---|---|---|---|---|---|---|---|---|---|---|---|---|---|---|---|---|
| **Entry Mode** | direct | direct | direct | direct | orbit | orbit | direct | direct | direct | Direct | direct | direct | direct | direct | direct | direct |
| **Entry Velocity** | | | | | 4.61 | 4.74 | - | 7.26 | 5.75 | 4.9 | 6.91 | 5.63 | | 5.59 | | |
| **Relative Entry Velocity** | 6.0 | 5.7 | 5.6 | 1.2 | 4.42 | 4.48 | - | 7.48 | 4.9 | 4.6 | 6.80 | 5.40 | 6.9 | 5.67 | 5.83 | 6.3 |
| **Peak Heat Velocity** | | | | | 4.02 | 4.0 | - | 6.61 | | - | | 4.70 | 5.94 | | | |
| **Aeroshell** | 120° break cone | 120° break cone | 120° break cone | 120° break cone | 70°Sphere Cone | 70°Sphere Cone | - | 70°Sphere Cone | Blunt Ended Cone | Blunt Ended Cone | 70°Sphere Cone | 60° Sphere Cone | 45° Sphere Cone | 70°Sphere Cone | 70°Sphere Cone | 70°Sphere Cone |
| **Base Area** | | | | | 9.65 | 9.65 | - | 5.52 | | | | 0.62 | | | | |
| **Ballistic Coefficient** | | | | | 64 | 64 | - | 62.3 | | | 65 | 69.9 | 36.2 | 70 | 82 | |
| **Entry Mass** | 1210 | 1210 | 635 | 635 | 992 | 992 | 2600 | 584 | 120.5 | 45 | 487 | 72.7 | | 603 | 577 | 608 |
| **Mass** | 358 | 358 | 635 | 635 | 590 | 590 | 6220 | 361 | 3159 | 88 | 583 | 33.2 | 2.4 | | 577 | 727 |
| **Aeroshell Diameter** | 3.2 | 3.2 | 3.2 | 3.2 | 3.5 | 3.5 | - | 2.65 | 1 | 0.29 | 2.4 | 0.93 | | 2.65 | 2.4 | 6.1 |
| **Parachute Diameter** | | | | | 16 | 16 | - | 12.5 | - | | | 10.4 | | 11.5 | 12 | |
| **Parachute Drag** | | | | | 0.67 | 0.67 | - | 0.4 | - | | | 0.92 | | 0.62 | | |
| **Lift/ Drag ratio** | | | | | 0.18 | 0.18 | - | 0 | - | | | 2° | | 0 | 0 | |
| **Vertical Velocity** | | | | | 2.4 | 2.4 | - | 12.5 | 0.02 | 0.075 | 2.4 | | | 2.4 | | 2.3 |
| **Landing Legs** | Ring | Ring | Ring | Ring | 3 | 4 | - | 0 | Ring | | 3 | 0 | 0 | 3 | 0 | 3 |
| **Touchdown Mass** | 358 | 358 | 635 | 635 | 590 | 590 | - | 360 | 30.6 | 4.5 | 423.6 | 33.2 | 2.4 | 364 | 280 | 360 |
| **Touchdown Velocity** | | | 0.06 | | 2.4 | 2.4 | - | 12.5 | 0.02 | 0.075 | 2.4 | | 0.17 | 2.4 | 0.15 | |
| **Attenuator Material** | Foam Plastic | Foam Plastic | Foam Plastic | Foam Plastic | Aluminium honeycomb | Aluminium honeycomb | Foam Plastic | Air-bags | Air-bags | Inflatable Ballute | Aluminium honeycomb | Air-bags | Hard Lander | Crushable | Crushable – Al Sanwich with carbon fiber | Crushable |
| **Landing Site** | 45°S 47°E | 45°S 202°E | 23.90°S 19.42°W | - | 22.27°N 47.95°W | 47.64°N 225.71°W | - | 19°7N 33°13W | 41°31N 153°77W | - | 76°S 195°W | 11.53°N 90.43°E | 73°S 210°W | 68.22°N 125.7°W | 6.208°W 2.052°S | 4.5°N 135.0°E |
| **MOLA** | | | | | -3.5 | -3.5 | - | -2.5 | | | -3.0 | | | -3.5 | 1.45 | -2.5 |

**Table-4 Lander Proposals under Study:**

| S.No | Name | Country | Agency | Proposed Launch | Type |
|---|---|---|---|---|---|
| **01.** | Icebreaker Life | United States | NASA | 2026 | Lander |
| **02.** | Martian Moon Exploration | Japan | JAXA | 2024 | Lander and Sample Return |
| **03.** | Phootprint | Europe | ESA | 2024 | Lander and Ascent Stage |
| **04.** | Fobos-Grunt (Repeat) | Russia | ROSCOSMOS | 2024 | Lander and Ascent Stage |
| **05.** | Mars-Grunt | Russia | ROSCOSMOS | 2024 | Orbiter and Lander |
| **06.** | BOLD | United States | NASA | 2020 | 6 – Impact Landers |

## VIII. Conclusion

In contrast to the accidental event of past crashed landers and the demand for future Mars exploration landing missions, a novel approach to landing architecture is discussed along with the structural configuration of the lander is developed. The geometric and aerodynamically approaches were described for faster aerocapture and successful landing by lowering the ballistic coefficient. Additionally, the EDL summary of past landers is compared (table-4) and the new formulation was done for shock attenuation. For this novel EDL system, vertical landing is preferred rather than horizontal landing due time limiting factor, and to effectuate a perfect landing, the potential to inflatable aerodynamic decelerators and orbital entry is approached. In upcoming years, some of the future proposed mission which was under study is discussed. The landing challenges for Mars landers were studied and some modifications were made to EDL sequences. This paper relies on the conceptual design of landing legs with good shock absorption capability. The simulation method for this model and prototype design will be studied further.

## Acknowledgments

The authors would like to thank Dr. Ahmed Sahinoz from Sabanci University for providing Solidworks software packages. Also, we would like to extend our sincere thanks to K Priyanga, an M.Sc. Graduate from our department for aiding in deriving mathematical formulation for the lander.

## Conflict of Interest

The authors have no conflict of interest to report.

## Funding

No external funding was received to support this study.